\documentclass[baaa]{baaa}

\usepackage[pdftex]{hyperref}
\usepackage{subfigure}
\usepackage{natbib}
\usepackage{helvet,soul}
\usepackage[font=small]{caption}

\begin{document}


\journalvol{60}
\journalyear{2018}
\journaleditors{P. Benaglia, A.C. Rovero, R. Gamen \& M. Lares}


\contriblanguage{1}


\contribtype{1}

\thematicarea{3}

\title{Effects of neutrino mixing upon electron fraction in core collapse supernovae}


\titlerunning{Macro BAAA60 con instrucciones de estilo}


\author{M.M. Saez\inst{1,3}, M.E. Mosquera\inst{1,2,3}, O. Civitarese \inst{2,3}} 
\authorrunning{Saez et al.}


\contact{msaez@fcaglp.unlp.edu.ar}

\institute{ Facultad de Ciencias Astron\'omicas y Geof\'{\i}sicas (UNLP) \and 
  Departamento de F\'{\i}sica (UNLP) \and CONICET }


\resumen{La inclusi\'on de neutrinos masivos afecta a las secciones eficaces involucradas en las cadenas de formación de núcleos pesados, alterando sus abundancias.
Los procesos rápidos de captura neutrónica ( proceso-r) suelen asociarse con eventos explosivos como las supernovas por colapso del núcleo. En este trabajo estudiamos los efectos de la incorporación de las  masas de los neutrinos, el impacto de la inclusión de un sabor estéril y las consecuencias de su oscilación con neutrinos activos sobre la tasa de neutrones libres, los flujos de neutrinos, la densidad bariónica  y la fracción electrónica del material. Hemos considerado dos propuestas diferentes para la funci\'on de distribuci\'on inicial de los neutrinos y distintas combinaciones de par\'ametros de mezcla (incluyendo $\theta_{34}\ne0$). En los cálculos trabajamos con el formalismo de matrices densidad, incluyendo los efectos de la oscilación, las interacciones con la materia y las interacciones neutrino-neutrino. Encontramos que las interacciones neutrino-materia y neutrino-neutrino modifican la fracci\'on electr\'onica, afectando la probabilidad de ocurrencia y el desenlace del proceso r.}

\abstract{
The inclusion of massive neutrinos affects the cross sections involved in the formation of heavy nuclei, modifying their abundances.
Rapid neutron capture processes (r-process) are often associated with explosive events such as core-collapse supernovae. In this work we study the effects of active and sterile neutrino oscillations and interactions, upon the calculation of  neutrino fluxes, the baryonic density and the electron fraction of the material. We have considered two different  initial distribution functions of the neutrinos and different combinations of mixing parameters (including $\theta_{34}\ne0$). We use the formalism of density matrices for the calculations and included the effects of neutrino oscillations, interactions with matter and self-neutrino interactions. We found that the interactions of the neutrinos with matter and with themselves  change the electron fraction, affecting the onset of the r-process. 
}



\maketitle

\section{Introduction}
\label{S_intro}

The results of detectors of  solar, atmospheric and reactors neutrinos such as LSND (Liquid Scintillator Neutrino Detector), SK (Kamiokande), SNO (Sudbury Neutrino Observatory) have provided evidence of neutrino oscillations caused by non-zero neutrino masses.

The results published in recent years by LSND and MiniBooNE (Mini Booster Neutrino Experiment) have established limits for the existence of other extra type of neutrino: the sterile \citep{athanassopoulos96,aguilar13}.With the motivation of these experimental outcome the inclusion of sterile neutrinos in different astrophysical scenarios are being analysed. 


 In the context of supernovae (SN), the active neutrino flux might suffer conversions to the sterile flavor causing a lower flow of electron neutrinos \citep{molinari03}. The effects of neutrino oscillations in supernova explosions have been studied by several authors \citep{balasi15,fetter03,balantekin04,tamborra12,wu15,janka12,woosley94,qian03}.

The rapid neutron-capture process (r-process) is responsible for the formation of heavy nuclei. This process requires a neutron-rich environment, i.e. an electron fraction per baryon ($Y_\textrm{e}$) lower than $0.5$, sufficiently large entropy, and sufficiently fast time scales, indicating that r-process sites are associated with explosive phenomena. In particular, the neutrino driven matter outflow (generated in later times in the SN, bounce time post $ t_ \textrm{{pb}} \sim 10 $ sec)  is a candidate site for the formation of elements beyond iron by the r-process \citep{qian03}. The neutron richness of the wind, is determined by the reactions $\nu_e+n\rightarrow p +e^-$ and $\bar{\nu_\textrm{e}}+p \rightarrow n + e^+$ \citep{qian96}.

In this work, we study the impact of neutrino oscillations over the electron fraction in the late neutrino-driven wind epoch. For this purpose, we compute the neutrino number densities, fluxes and luminosities for different cases.

\section{Description of the environment}

The  neutrino driven-wind is generated after the rebound of the collapsing star. For large radius we can obtain the baryon mass-density as \citep{balantekin04}
\begin{eqnarray}\label{rhob}
\rho_\mathrm{b} &\simeq &38 \frac{2}{11} g_\mathrm{s} \frac{M_\mathrm{{NS}}^3}{S_\mathrm{{100}}^4r_\mathrm{7}^3} \, ,
\end{eqnarray}
expressed in units of $10^3 {\rm gr\, cm}^{-3}$. $S_\textrm{100}$ is the entropy per baryon in units of $100\, k_\textrm{B}$ and $r_7$ is the distance to the center of the star in units of $10^7 \, {\rm cm}$, $M_\textrm{NS}$ is the mass of the proto-neutron star and $ g_\textrm{s}$ is he number of degrees of freedom \citep{balantekin04}.

Since different values of the entropy indicate different stages of the SN evolution, we have used a fixed value of $S_\textrm{100}=1.5$, representing the late cooling phase of the neutrino-driven wind \citep{janka07}.
{Neutrino fluxes can be obtained, after integration on solid-angles as \citep{balantekin04}
\begin{eqnarray}\label{phi}
\frac{d\phi_\nu}{dE_\nu}&=&\frac{c}{8\pi^3(\hbar c)^3}\frac{R^2_\nu}{r^2}f_\nu(E_\nu) \, ,
\end{eqnarray}
where $R_\nu$ is the radius of the neutrino-sphere and $f_{\nu}(E_{\nu})$ is the amount of neutrinos for each flavor for an specific radius. }
Weak reactions modify the amount of neutrons. The rate  can be computed as \citep{tamborra12}
\begin{eqnarray}\label{rate}
\lambda_\nu&=&\int \sigma_\nu(E_\nu)\frac{d\phi_\nu}{dE_\nu }dE_\nu \, ,
\end{eqnarray}
where the cross section, in units of ${\rm cm}^2$, are $\sigma_{\nu}(E_{\nu})=9.6\times10^{-44}\left(\frac{E_{\nu}\pm\Delta m_\mathrm{np}}{MeV}\right)$.
In the last expressions the $+$ is for neutrinos and the $-$ for anti-neutrinos, $\Delta m_\textrm{np}= 1.293 \, \rm{MeV}$ is the neutron-to-proton mass-difference. 

If the plasma reaches a weak equilibrium stage, the electron fraction of the material $Y_\textrm{e}$ can be written as \citep{mclaughlin96}
\begin{eqnarray}\label{ye}
Y_\mathrm{e}&=&\frac{\lambda_\mathrm{n}}{\lambda_\mathrm{n}+\lambda_\mathrm{p}}+\frac{1}{2}\frac{\lambda_\mathrm{p}-\lambda_\mathrm{n}}{\lambda_\mathrm{p}+\lambda_\mathrm{n}} X_\alpha \, .
\end{eqnarray}
where $\lambda_\mathrm{p}=\lambda_{\bar{\nu}_\mathrm{e}}+\lambda_{\mathrm{e}^-}$, $\lambda_\mathrm{n}=\lambda_{\nu_\mathrm{e}}+\lambda_{\mathrm{e}^+}$ and $X_\alpha$ is the mass fraction of $\alpha$ particles.
In deriving the above equations we have taken $X_{\alpha}$ as a time independent quantity. 


\section{Neutrino Oscillations and interactions}\label{Sec:3}

 Calling $\rho$ ($\bar{\rho}$) to the neutrino (antineutrino)-distribution
function in its matrix form and $\mathcal{H}$ ($\bar{\mathcal{H}}$) the neutrino (anti-neutrino) Hamiltonian in the flavor basis, the equations that give the neutrino distribution function as a function of the radius are \citep{balantekin04,tamborra12}
\begin{eqnarray}
i\frac{\partial \rho}{\partial r}=\left[\mathcal{H},\rho\right]\nonumber \, ,\hspace{1cm} i\frac{\partial \bar{\rho}}{\partial r}=\left[\bar{\mathcal{H}},\bar{\rho}\right]\, .
\end{eqnarray}

Where $\mathcal{H}=\mathcal{H}^{\textrm{vac}}+\mathcal{H}^{\textrm{m}}+\mathcal{H}^{\nu-\nu}$.
$\mathcal{H}^{\textrm{vac}}$ describes the neutrino oscillations in vacuum, $\mathcal{H}^{\textrm{m}}$ represents the neutrino-matter interactions and $\mathcal{H}^{\nu-\nu}$ takes into account the neutrino-neutrino interactions. In this treatment of $\nu-\nu$ interactions we assume the single-angle approximation in which all neutrinos feel the same neutrino-neutrino refractive effect \citep{duan06,duan10}.

\subsection{Mixing of two active-neutrinos}

In this case, since the tau and muon-neutrino fluxes are similar in a SN, one can assume a combination of both 
type of neutrinos, the so-called $x$-neutrino \citep{tamborra12}, which can mix with the electron-neutrino through
a mixing angle $\theta_{13}$. The neutrino Hamiltonian in the flavor basis reads
\begin{eqnarray}
\hspace{-0.5cm}\mathcal{H}^{\mathrm{vac}}&=\left(pc+\frac{m_1^2 c^3}{2p}\right)\mathcal I_{2\times2}
+\frac{\Delta m^2_{13} c^3}{2p}\left(
\begin{array}{cc}
s_{13}^2 & c_{13}  s_{13}  \\
 c_{13}  s_{13} & c_{13}^2
\end{array}
\right)\, \,\,\,\,
\end{eqnarray}
where $p$ is the momentum, $m_i$ stands for the neutrino mass of the eigenstate $i$ and we have used the notation $c_{ij}=\cos(\theta_{ij})$, $s_{ij}=\sin(\theta_{ij})$ and $
\Delta m^2_{13}=m_3^2-m_1^2$.
The neutrino-electron and neutrino-neutron interactions are described by  the Hamiltonian
\begin{eqnarray}
\mathcal{H}^\mathrm{m}&=&\frac{\sqrt{2}}{2}G_\mathrm{f} N_\mathrm{b} \left(
\begin{array}{cc}
3 Y_\mathrm{e}-1&0\\
0&Y_\mathrm{e}-1
\end{array}
\right) \, ,
\end{eqnarray}
where $N_\mathrm{b}$ is the baryon density {and we have obtained it from eq. \ref{rhob}} \citep{tamborra12}. 

The neutrino-neutrino Hamiltonian is
\begin{eqnarray}
\mathcal{H}^{\nu-\nu}&=&\sqrt{2}G_\mathrm{f}\left[\Delta N_\mathrm{e}\left(
\begin{array}{cc}
2&0\\
0&1
\end{array}
\right)
+ \Delta N_\mathrm{x}\left(
\begin{array}{cc}
1&0\\
0&2
\end{array}
\right)\right], \,\,\,\,\,\,\,\,\,\,
\end{eqnarray}
where $G_\textrm{f}$ is the Fermi constant and $\Delta N_i$ is the difference between the density of the $i$-flavour neutrino and antineutrino.


\subsection{Active-sterile mixing, $2+1$ scheme}

In this case we have have considered that the light neutrino can oscillate with a sterile neutrino of mass $m_4$. The Hamiltonian $\mathcal{H}^{\textrm{vac}}$ reads 
\begin{eqnarray}
&&\hspace{-0.6cm}\mathcal{H}^{\mathrm{vac}}=\left(pc+\frac{m_1^2c^3}{2p}\right) \mathcal I_{3\times3}\nonumber \\&&\hspace{-0.6cm}
+\frac{\Delta m^2_{14} c^3}{2p}\left(
\begin{array}{ccc}
s^2_{14} & 0& c_{14} s_{14}\\
0&0&0\\
c_{14} s_{14}& 0&c^2_{14}
\end{array}
\right)
 \\ &&
\hspace{-0.6cm}+\frac{\Delta m^2_{13} c^3}{2p}\left(
\begin{array}{ccc}
c^2_{14}s^2_{13} & c_{14} c_{13}s_{13} &-c_{14} s_{14}s^2_{13}\\
c_{14} c_{13}s_{13} &c^2_{13}& -c_{13}s_{13}s_{14}\\
-c_{14} s_{14} s^2_{13}& -s_{14} c_{13}s_{13} &s^2_{14}s^2_{13}
\end{array}
\right)\nonumber
 .
\end{eqnarray}

The neutrino-matter interaction can be computed as
\begin{eqnarray}
\mathcal{H}^\mathrm{m}&=&\frac{\sqrt{2}}{2}G_\mathrm{f} N_\mathrm{b} \left(
\begin{array}{ccc}
3 Y_\mathrm{e}-1 & 0 & 0\\
0 & Y_\mathrm{e}-1 & 0\\
0 & 0 & 0
\end{array}
\right) \, .
\end{eqnarray}
Finally, the neutrino-neutrino interaction is written
\begin{eqnarray}
\hspace{-0.6cm}\mathcal{H}^{\nu-\nu}&=&\sqrt{2} G_\mathrm{f}\left[\Delta N_\mathrm{e}\left(
\begin{array}{ccc}
2 & 0 & 0 \\
0 & 1 & 0\\
0 & 0 & 0
\end{array}
\right)+\Delta N_\mathrm{x}\left(
\begin{array}{ccc}
1 & 0 & 0 \\
0 & 2 & 0\\
0 & 0 & 0
\end{array}
\right)\right].\,\,\,\,\,\,\,\,\,\,\,\,
\end{eqnarray}

\section{Results}
We have solved the evolution equations {of sec. \ref{Sec:3} coupled to Eqs. \ref{rhob}-\ref{ye}} to compute the amount of neutrinos for each flavor and $Y_e$ as a function of the radius for the late cooling phase of the neutrino-driven wind $t_\textrm{{pb}}\sim 10 \, \rm{sec}$. For this purpose, we have calculated fluxes, reaction rates and electronic fraction in parallel. To solve the coupled differential equations we have adopted  values for the neutrino mixing parameters given in the literature \citep{meregaglia16,minakata04}. As initial condition we have taken  at the neutrino sphere radius $R_\nu=10 \, {\rm Km}$  two different distribution functions to characterize the neutrinos, namely a Fermi-Dirac (FD) distribution with the mean-energies were extracted from Refs. \citep{qian03,balantekin04} and  a power-law distribution \citep{keil03,tamborra12,pllumbi15}. 
 Also we have performed the calculation assuming different constant values for $X_\alpha$.

\begin{figure}[!h]

  \centering
  \includegraphics[width=0.50\textwidth]{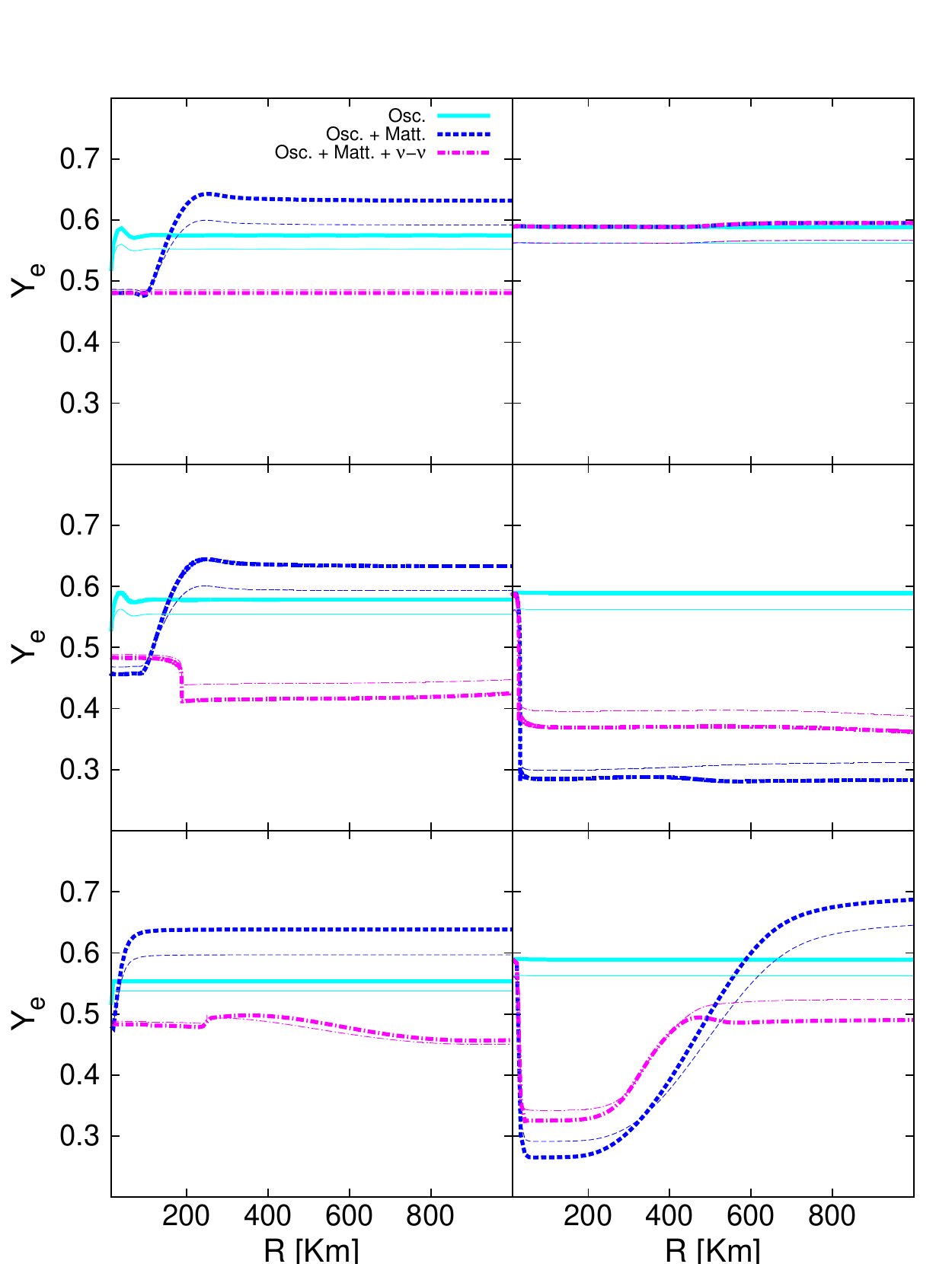}
  \caption{$Y_\textrm{e}$ as a function of the radius for the late cooling time. The columns indicates de different distribution functions used as initial condition: Fermi-Dirac (first column) and the power-law distribution (second column). The rows represent the different oscillation schemes: active-active neutrino oscillations (first row), active-sterile mixing with $\theta_{34}=0$ (second row) or $\theta_{34}\neq0$ (third row). Thicker and thinier lines represents $X_\alpha=0$ and $X_\alpha=0.3$ respectively.
}
  \label{fig:ye}
\end{figure}

In fig. \ref{fig:ye} we present the electron fraction, as a function
of the radius for $X_\alpha=0$ (thicker lines) and for $X_\alpha=0.3$ (thinner lines), including different interactions  and for different oscillation schemes. The first column of the figure was obtained using a Fermi Dirac distribution as initial condition, meanwhile the second column was computed using a power-law distribution function. We have use a normal hierarchy, $\Delta m_{13}^2=2\times 10^{-3} \, {\rm eV}^2$, $\Delta m_{14}^2=2\,{\rm eV}^2$, $\sin^2 2\theta_{13}=0.09$, 
and $\sin^2 2\theta_{14}=0.16$. The first row corresponds to the active-active scenario, the second and third rows represent the results in the $2+1$ scheme for the case of only one active-sterile mixing angle ($\theta_{14}$) and for two active-sterile mixing angles ($\theta_{14}$ and $\theta_{34}$) respectively \citep{collin16}. 

 As one can see the electrons suffer a depletion if the neutrino-matter interactions are turn on. If we considered also the neutrino-
neutrino interaction, the depletion is reduced, however the value of $Y_e$ is lower than the obtained without any interaction. 


\section{Conclusions}
In this work we have studied the impact of the inclusion of massive  neutrinos and sterile neutrinos upon the physical conditions required for the success of the r-process in a supernova environment. We have solved the coupled equations to calculate the electron-fraction in the stellar interior as a function of the mixing parameters. We have found that the electron abundance is sensitive to the inclusion of sterile neutrinos, and that it depends on the neutrino interactions considered, the set of oscillation parameters and the initial distribution function. 
 As general features we can mention that the inclusion of the sterile neutrino have an important effect upon $Y_\textrm{e}$, since it can be drastically reduced. The scenario of active-active oscillations is the most unfavourable to achieve $ Y_\textrm{e} \leq 0.5 $ (result in accordance with \cite{tamborra12,wu15}).
The fact that $ \theta_ {34}$ might not be null, has an important effect on $Y_\textrm{e}$.
The determination of the active-sterile neutrino oscillation parameters, the search for an appropriate description for the initial neutrino fluxes and a good modelling of the interactions involved, are relevant to understand and estimate the viability of the r-process


\begin{acknowledgement}
The work was financed by the National Council of Scientific and Technological Research of
Argentina (CONICET). The O.C. and M.E.M. authors are members of the Scientific Research Career of the CONICET.
\end{acknowledgement}


\bibliographystyle{baaa}
\small
\bibliography{biblio-saez}
 
\end{document}